\documentclass[twocolumn]{aastex62}

\submitjournal{The Astrophysical Journal}

\shorttitle{PIC simulation of heat flux instability}
\shortauthors{Kuzichev et al.}

\begin{document}

\title{Nonlinear evolution of the whistler heat flux instability}

\correspondingauthor{Ilya Kuzichev}
\email{ilya.kuzichev@njit.edu }

\author[0000-0001-6815-1591]{Ilya V. Kuzichev}
\affil{New Jersey Institute of Technology, Newark, USA}
\affil{Space Research Institute of Russian Academy of Sciences, Moscow, Russia}

\author[0000-0002-4974-4786]{Ivan Y. Vasko}
\affil{Space Sciences Laboratory, University of California, Berkeley, CA 94720-7450, USA}
\affil{Space Research Institute of Russian Academy of Sciences, Moscow, Russia}

\author[0000-0001-8883-1746]{Angel Rualdo Soto-Chavez}
\affil{New Jersey Institute of Technology, Newark, USA}

\author[0000-0002-3354-486X]{Yuguang Tong}
\affil{Space Sciences Laboratory, University of California, Berkeley, CA 94720-7450, USA}
\affil{Physics Department, University of California, Berkeley, CA 94720-7300, USA}

\author[0000-0001-8823-4474]{Anton V. Artemyev}
\affil{Institute of Geophysics and Planetary Physics, University of California, Los Angeles, USA}
\affil{Space Research Institute of Russian Academy of Sciences, Moscow, Russia}

\author[0000-0002-1989-3596]{Stuart D. Bale}
\affil{Physics Department, University of California, Berkeley, CA 94720-7300, USA}
\affil{Space Sciences Laboratory, University of California, Berkeley, CA 94720-7450, USA}

\author[0000-0001-9179-9054]{Anatoly Spitkovsky}
\affil{Department of Astrophysical Sciences, Princeton University, New Jersey, USA}




\begin{abstract}
We use the one-dimensional TRISTAN-MP particle-in-cell code to model the nonlinear evolution of the whistler heat flux instability that was proposed by \cite{Gary1999b} and \cite{Gary2000a} to regulate the electron heat flux in the solar wind and astrophysical plasmas. 
The simulations are initialized with electron velocity distribution functions typical for the solar wind. We perform a set of simulations at various initial values of the electron heat flux and $\beta_{e}$. The simulations show that parallel whistler waves produced by the whistler heat flux instability saturate at amplitudes consistent with the spacecraft measurements. The simulations also reproduce the correlations of the saturated whistler wave amplitude with the electron heat flux and $\beta_{e}$ revealed in the spacecraft measurements. The major result is that parallel whistler waves produced by the whistler heat flux instability do not significantly suppress the electron heat flux. The presented simulations indicate that coherent parallel whistler waves observed in the solar wind are unlikely to regulate the heat flux of solar wind electrons. \end{abstract}



\section{Introduction}

The electron heat flux in collisionless or weakly collisional plasma is generally not described by the collisional Spitzer-H{\"a}rm law \citep{Spitzer53}. The heat flux suppression below the collisional value was demonstrated by direct in-situ measurements in the solar wind \citep{Feldman75,Scime94,Crooker03,Bale13} and remote observations of the temperature distribution of a hot gas in galaxy clusters \citep{Cowie77,Meiksin86,Fabian94,Zakamska&Narayan03,Wagh14,Fang18}. The Spitzer-H{\"a}rm law was hypothesized to be inadequate for describing the heat conduction in the solar corona \citep[][]{Scudder92,Landi01,Dorelli03} and that was recently confirmed in the analysis of coronal loop oscillations \citep[][]{Wang15}. The observations in the solar wind were interpreted in terms of the heat flux regulation by wave-particle interactions \citep{Feldman75,Feldman76,Gary77,Scime94,Crooker03}, but the wave activity potentially regulating the electron heat flux is still under debate \citep[e.g.,][]{Gary1999b,Scime01,Pagel07,Roberg-Clark:2016,Tong18:arxiv,Roberg-Clark18:apj,Roberg18:prl,komarov_2018,Vasko19,Verscharen19}. The alternative view is that wave-particle interactions may be not necessary to explain the observed heat flux values in the solar wind \citep[][]{Landi2012,Landi14,Horaites15}.

The mechanism of the electron heat flux regulation is intimately related to kinetic features of the electron velocity distribution function (VDF). The electron VDF in a slow solar wind is often adequately described by a combination of bi-Maxwellian thermal dense core and suprathermal tenuous halo populations \citep[][]{Feldman75,Feldman76,Maksimovic97,Tong19}. In the plasma rest frame, the core and halo populations are streaming along the background magnetic field, the net electron current is approximately zero, while the electron heat flux is anti-sunward and carried predominantly by halo electrons \citep[][]{Feldman75,Feldman76,Scime94,Tong19}. In what follows we focus on a slow solar wind and do not consider effects of the strahl population (anti-sunward beam-like population) frequently observed in a fast solar wind and carrying a significant portion of the electron heat flux \citep[e.g.,][]{Rosenbauer77,Pilipp87,Stverak09}.

At sufficiently large heat flux values electrons are capable of generating whistler waves propagating quasi-parallel to the background magnetic field \citep{Gary75,Gary1994a}. Spacecraft measurements of the electron heat flux values below a bound dependent on $\beta_{e}$ (electron beta parameter) were interpreted in terms of the heat flux regulation by this so-called whistler heat flux instability (WHFI) \citep{Feldman76,Gary77,Gary1999b}. The argument behind that hypothesis was that the heat flux bound dependent on $\beta_{e}$ is similar to the marginal stability threshold of the WHFI. \cite{Gary2000a} extrapolated the hypothesis and proposed that the WHFI may regulate the electron heat flux in high-$\beta_{e}$ astrophysical plasma. \cite{Pistinner1998a} called into question that hypothesis, but no detailed analysis of the nonlinear stage of the classical WHFI was performed to support their arguments. \cite{Scime01} questioned that the WHFI regulates the electron heat flux in the solar wind based on Ulysses measurements.

The simultaneous wave and particle measurements have recently shown that, though intermittently, whistler waves are indeed present in the solar wind and they are predominantly quasi-parallel that is consistent with the WHFI scenario \citep{Lacombe14,Stansby16,Kajdic16,Tong19,Tong19b}. \cite{Tong19} have demonstrated for several events that the whistler waves were indeed generated locally by the WHFI. The extensive statistical analysis by \cite{Tong19b} has shown that in the solar wind whistler waves have amplitudes typically less than a few hundredths of the background magnetic field. The amplitude is positively correlated with $\beta_{e}$ and generally with the electron heat flux. Though the WHFI is indeed operating in the solar wind, there has been no analysis that would establish whistler wave properties in the nonlinear stage of the WHFI and clarify whistler wave effects on the electron heat flux evolution.

In this paper, we present Particle-In-Cell (PIC) simulations of the WHFI using the fully relativistic massively parallelized TRISTAN-MP code \citep{Spitkovsky08,Park2015}. The simulations are restricted to parallel whistler waves, because they have the largest growth rates according to the linear theory of the WHFI \citep{Gary75,Gary1994a}, and initialized with proton and electron VDFs typical for the solar wind. The simulations demonstrate that in a uniform plasma parallel whistler waves produced by the WHFI are incapable of regulating the electron heat flux. The whistler waves are shown to saturate at amplitudes consistent with the recent spacecraft measurements. The paper is organized as follows. We summarize the results of the WHFI linear theory in Section \ref{sec1}, then present results of the PIC simulations in Section \ref{sec2}, discuss and summarize our conclusions in Sections \ref{sec3} and \ref{sec4}.

\section{WHFI linear theory \label{sec1}}

In the linear theory of the original WHFI \citep{Gary75}, the electron VDF consists of Maxwellian core and halo populations streaming along the background magnetic field, $f_e(v_{||},v_{\perp})=f_{c}(v_{||},v_{\perp})+f_{h}(v_{||},v_{\perp})$, where
\begin{eqnarray*}
f_{\alpha}&=&n_{\alpha}\left(\frac{m_e}{2\pi T_{\alpha}}\right)^{3/2}\exp\left[-\frac{m_{e}\left(v_{||}-u_{\alpha}\right)^2+m_{e}v_{\perp}^2}{2T_{\alpha}}\right],
\end{eqnarray*}
and $n_{\alpha}$, $u_{\alpha}$ and $T_{\alpha}$ denote density, bulk velocity and temperature of the core ($\alpha = c$) and halo ($\alpha = h$) populations, $m_{e}$ is the electron mass. The electron current in the plasma rest frame is assumed to be zero, $n_{c}u_{c}+n_h u_{h}=0$. The electron heat flux associated with the core plus halo electron VDF, $q_{e}=-0.5\;n_c u_{c}\;(5 T_{h}+m_{e}u_{h}^2-5T_{c}-m_{e}u_{c}^2)$, is often normalized to the free-streaming heat flux value $q_0=1.5\; n_e T_e (2T_e/m_{e})^{1/2}$, where $n_e=n_c+n_h$ is the total electron density, $T_{e}=(n_cT_{c}+n_hT_{h})/n_{e}$ is typical macroscopic electron temperature \citep[e.g.,][]{Gary1999b}. The normalized electron heat flux $q_{e}/q_0$ depends on $n_{c}/n_e$, $T_{h}/T_{c}$ and $u_{c}\sqrt{m_e/T_{c}}$. At realistic parameters in the solar wind, thermal protons do not interact resonantly with whistler waves unstable to the WHFI \citep[e.g.,][]{Gary1994a,Gary2000a}, therefore the contribution of protons to the whistler wave linear dispersion relation is considered in the frame of the cold fluid approach.

\begin{figure*}
    \centering
    \includegraphics[width=0.5\linewidth]{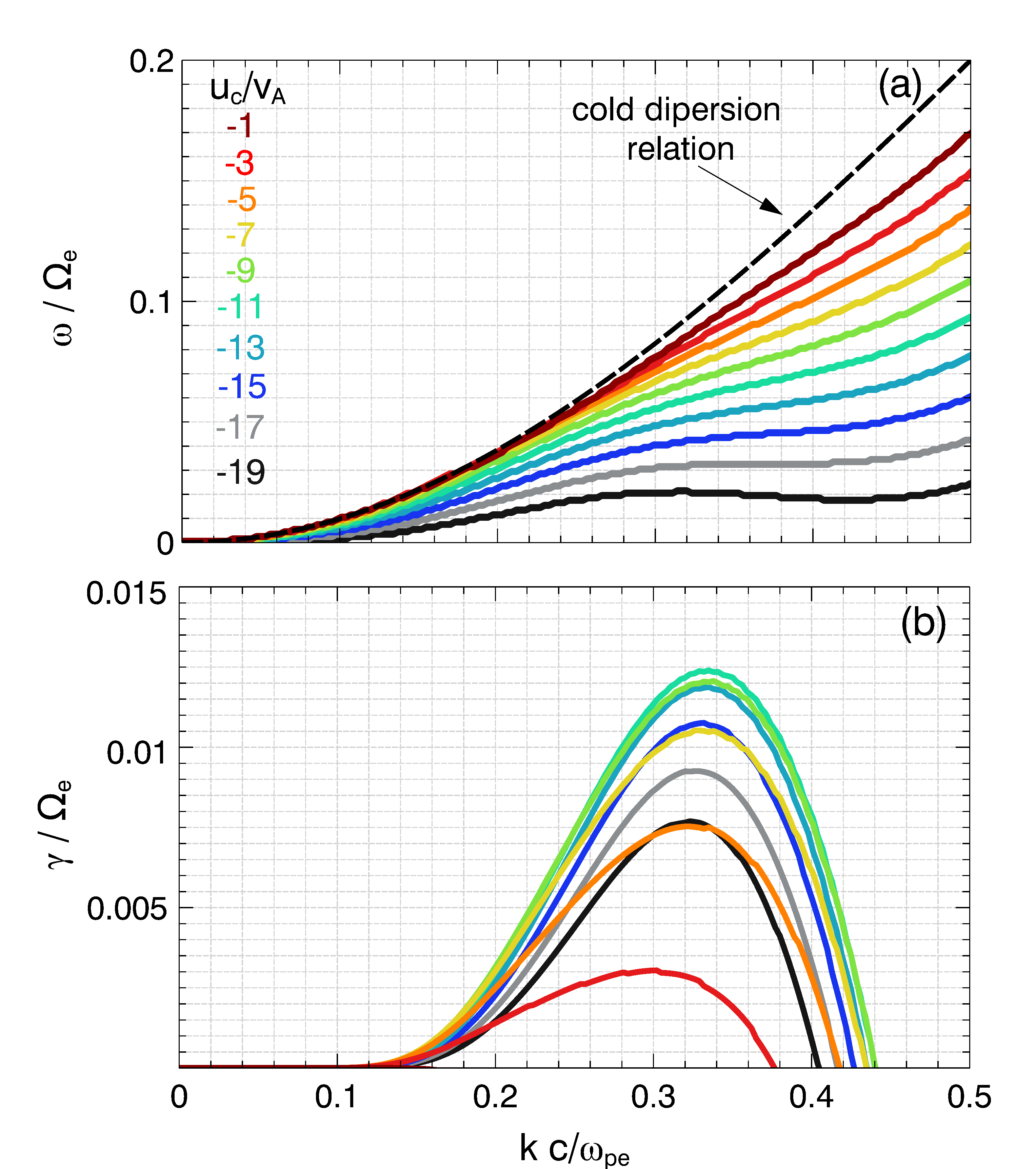}
    \caption{Summary of the linear stability analysis of the whistler heat flux instability in a plasma with $n_{c}/n_{e}=0.85$, $T_{h}/T_{c}=10$ and $\beta_{c}=1$. Panels (a) and (b) show dispersion curves and linear growth rates of parallel whistler waves computed at various $u_{c}/v_{A}$, where $u_{c}$ is the bulk velocity of the core population, $v_{A}=B_0/(4\pi n_e m_p)^{1/2}$ is the Alfv\'en velocity. The whistler wave frequency and growth rate are normalized to the electron cyclotron frequency $\Omega_{e}$, while the wavenumber is normalized to the electron inertial length $c/\omega_{pe}$. Panels (a) and (b) show that the fastest growing whistler waves have wavelengths of about 20 $c/\omega_{pe}$ and frequencies below $0.1\,\Omega_e$.}
    \label{fig1}
\end{figure*}

At $\omega_{pe}\gg \Omega_e$ the frequency and the linear growth rate of a parallel whistler wave are determined by the following dispersion relation \citep[e.g.,][]{Mikhailovskii74}  
\begin{eqnarray*}
\frac{k^2c^2+\omega_{pi}^2}{\omega^2}\approx \frac{\omega_{pe}^2}{\omega^2}\sum_{\alpha=c,h}\frac{n_\alpha}{n_e}\;\frac{(\omega-ku_{\alpha})}{kv_{\alpha}}\;Z\left(\frac{\omega-ku_{\alpha}-\Omega_{e}}{kv_{\alpha}}\right)
\label{eq:1}
\end{eqnarray*}
where $\omega$ and $k$ are the whistler wave frequency and wavenumber, $\omega_{pe}$ and $\omega_{pi}$ are electron and proton plasma frequencies, $\Omega_{e}$ is the electron cyclotron frequency, $Z(\xi)=(2\pi)^{-1/2}\int_{-\infty}^{+\infty} dx\;e^{-x^2/2}/(x-\xi)$ is the plasma dispersion function, $v_{\alpha}=(T_{\alpha}/m_{e})^{1/2}$ denotes thermal velocity of the core and halo populations. We solve the dispersion relation in the limit of a weak instability via the standard techniques \citep[e.g.,][]{Mikhailovskii74}. The linear growth rate $\gamma$ normalized to $\Omega_{e}$ depends on $kc/\omega_{pe}$, $n_{c}/n_{e}$, $T_h/T_c$, $\beta_{c}=8\pi n_c T_c/B_0^2$ and $u_{c}/v_{A}$, where $v_{A}=B_0/(4\pi n_e m_p)^{1/2}$ is the Alfv\'en velocity, $B_0$ is the background magnetic field, $m_p$ is the proton mass. In what follows, we present results of the linear stability analysis of the WHFI for $n_{c}/n_{e}=0.85$ and $T_{h}=10\,T_{c}$ if not stated otherwise.

Figure \ref{fig1} presents results of the linear stability analysis at $\beta_{c}=1$ and various $u_{c}/v_{A}$ in the range from $-1$ to $-19$. Panel (a) shows that the whistler wave dispersion curves are dependent on $u_{c}/v_{A}$ and generally different from the whistler dispersion curve in a cold plasma, $\omega=\Omega_{e} k^2c^2/(k^2c^2+\omega_{pe}^2)$ \citep[e.g.,][]{Mikhailovskii74}. Panel (b) presents the whistler wave growth rates at various $u_{c}/v_{A}$ and demonstrates that the growth rates reach maximum values of about 0.01 $\Omega_{e}$ at $kc/\omega_{pe}\approx 0.3$. The wavelength of the fastest growing whistler waves is about 20 $c/\omega_{pe}$. We have performed similar growth rate computations at $\beta_{c}=0.4, 2$ and $3$ and determined parameters of the fastest growing whistler waves.

\begin{figure*}
    \centering
    \includegraphics[width=0.5\linewidth]{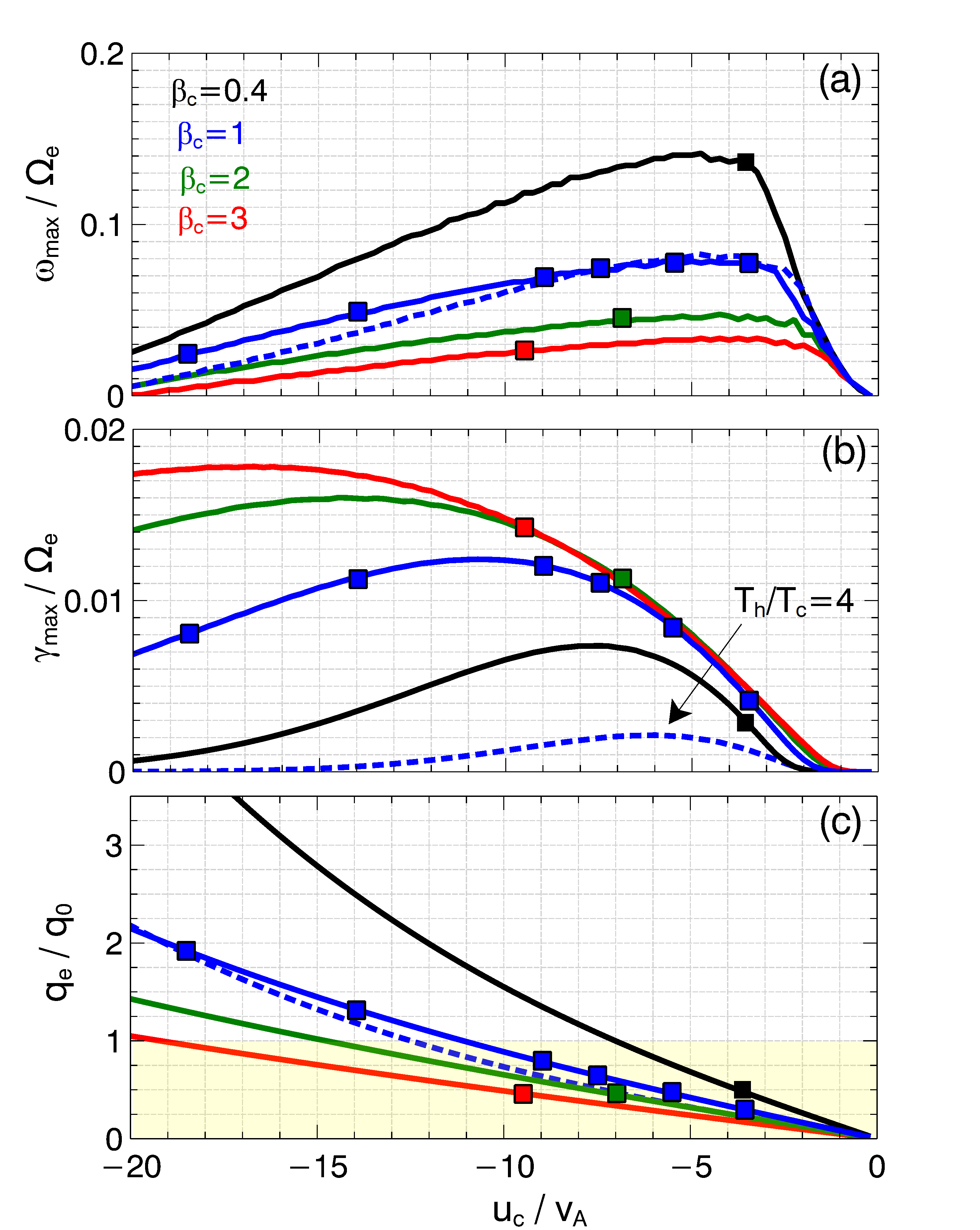}
    \caption{Summary of the linear stability analysis of the whistler heat flux instability in a plasma with $n_{c}/n_{e}=0.85$, $T_{h}/T_{c}=10$ (solid curves) and $n_{c}/n_{e}=0.85$, $T_{h}/T_{c}=4$ (dashed curve). Panels (a) and (b) present frequencies and growth rates of the fastest growing parallel whistler waves at various $u_{c}/v_{A}$ and $\beta_{c}$. Panel (c) presents the normalized electron heat flux $q_{e}/q_0$ at various $u_{c}/v_{A}$ and $\beta_{c}$. The squares correspond to initial conditions for the PIC simulations presented in Section \ref{sec2}. The shaded region in panel (c) indicates that in the realistic solar wind we generally have $q_{e}/q_0\lesssim 1$ \citep[e.g.,][]{Tong19b}}
    \label{fig2}
\end{figure*}

Figure \ref{fig2} presents parameters of the fastest growing whistler waves in dependence on $u_{c}/v_{A}$ at various $\beta_{c}$. Panel (a) shows that whistler waves at lower frequencies are preferentially unstable at larger $\beta_{c}$. Panel (b) shows that the growth rate of the fastest growing whistler wave is larger at larger $\beta_{c}$. Panel (c) presents the normalized heat flux $q_{e}/q_0$ as a function of $u_{c}/v_{A}$ for the further references. The normalized heat flux $q_{e}/q_0$ varies with $u_{c}/v_{A}$ and $\beta_{c}$, because it depends on $m_{e}u_{c}^2/T_{c} \sim u_{c}^2/v_{A}^2\beta_{c}$. Panels (b) and (c) show that at any given $\beta_{c}$, the growth rate of the fastest growing whistler wave is a non-monotonous function of $u_{c}/v_{A}$ or, equivalently, $q_e/q_0$ \citep[in accordance with][]{Gary85}. 

In addition to the linear theory results for $T_h/T_c = 10$, Figure \ref{fig2} also presents parameters of the fastest growing whistler waves in a plasma with $\beta_{c}=1$ and $T_{h}/T_{c}=4$. 
The comparison of the growth rates computed at $T_{h}/T_{c}=4$ and 10 shows that the smaller halo temperature results in more than five times smaller maximum growth rate. The specific values of $u_{c}/v_{A}$ and $q_{e}/q_0$ at which the growth rate reaches maximum depend on $T_{h}/T_{c}$ and $n_{c}/n_e$. For example, panels (b) and (c) show that at $T_{h}/T_{c}=10$, the growth rate reaches maximum at $u_{c}\approx -10\,v_{A}$ and $q_{e}/q_0\approx 0.95$, while at $T_{h}/T_{c}=4$, it reaches maximum at $u_{c}\approx -6\, v_{A}$ and $q_{e}/q_0\approx 0.5$.

\begin{table}[]
    \centering
    \begin{tabular}{|c|c|c|c|}
    \hline
    $\#$ & $-u_{c}/v_{A}$ & $q_{e}/q_0$  \\
    \hline
    1 &  3.5  & 0.3\\
    \hline
    2 &  5.5  & 0.45\\
    \hline
    3 &  7.5  & 0.65\\
    \hline
    4 &  9.0  & 0.8\\
    \hline
    5 &  14.0  & 1.3\\
    \hline
    6 &  18.5 & 1.9\\
    \hline
    \end{tabular}
    \caption{Parameters for the first set of simulations performed at $\beta_{c}=1$. Indicated are initial values of $u_{c}/v_{A}$ and $q_{e}/q_0$. In all these simulations $n_{c}/n_{e}=0.85$, $T_{h}/T_{c}=10$ and $\omega_{pe}/\Omega_e\approx 12.3$. In Figure \ref{fig2} the initial conditions for the first set of simulations are indicated by the blue squares.}
    \label{tab1}
\end{table}

\begin{table}[]
    \centering
    \begin{tabular}{|c|c|c|c|c|}
    \hline
    $\#$ &  $\beta_{c}$ & $-u_{c}/v_{A}$ & $\omega_{pe}/\Omega_e$  \\
    \hline
    1 &  0.4 & 3.5 & 7.8  \\
    \hline
    2 &  1   & 5.5 & 12.3  \\
    \hline
    3 &  2   & 7.8 & 17.3  \\
    \hline
    4 &  3   & 9.5 & 21.2   \\
    \hline
    \end{tabular}
    \caption{Parameters for the second set of simulations performed at the same initial heat flux value, $q_{e}/q_0=0.45$. Indicated are initial values of $\beta_{c}$, $u_{c}/v_{A}$ and $\omega_{pe}/\Omega_{e}$. In all simulations $n_{c}/n_{e}=0.85$ and $T_{h}/T_{c}=10$. In Figure \ref{fig2} the initial conditions for the second set of simulations are indicated by four squares (red, green, blue and black) at $q_{e}/q_0=0.45$.}
    \label{tab2}
\end{table}

\section{PIC simulations \label{sec2}}

We perform PIC simulations of the WHFI to establish the effect of the whistler waves on the electron heat flux evolution and determine the amplitude of whistler waves in the nonlinear stage. We use the TRISTAN-MP code \citep{Spitkovsky08, Park2015} with both ions and electrons treated as particles at realistic mass ratio. We restrict the analysis to 1D simulations with a single spatial coordinate $x$ along the background magnetic field, hence only strictly parallel whistler waves are modelled. The code is initialized with core and halo electron VDFs given by Maxwell-J{\"u}ttner distribution functions, which reduce to Maxwell distributions in a non-relativistic limit (that is the case in our simulations). The initial proton VDF is a Maxwell-J{\"u}ttner distribution function with the proton temperature equal to the core electron temperature \citep[that is realistic in the solar wind at 1 AU, see][]{Artemyev18:jgr}. Due to computational limitations, the simulations are performed at $\omega_{pe}/\Omega_e$ of the order of 10, while in the realistic solar wind this parameter is about ten times larger. At $\beta_{c}\sim 1$, this implies the temperature of the core population about 100 times larger than in the realistic solar wind (a few keV instead of 10 eV), because  $\omega_{pe}/\Omega_e=\beta_c^{1/2}(m_{e}c^2/2T_{c})^{1/2}$.  In fact, in all simulations $T_{c}=2$ keV, while the specific value of $\omega_{pe}/\Omega_e$ is determined by $\beta_{c}$. The length of the simulation box is $L_x = 1296\,c/\omega_{pe}$ that is much larger than the typical wavelength $20\,c/\omega_{pe}$ of the fastest growing whistler waves (Figure \ref{fig1}). The temporal and spatial integration steps are 0.09\,$\omega_{pe}^{-1}$ and 0.2 $c/\omega_{pe}$ that is adequate to reproduce the expected unstable whistler waves. The number of spatial cells is about 6480, the number of particles per cell is $4\cdot10^4$, giving the total number of particles $\approx 2.6\cdot 10^8$.

In all simulations, we assume $n_{c}/n_{e}=0.85$ that is realistic for the solar wind \citep[e.g.,][]{Maksimovic05,Tong19}. Typically in the solar wind $T_h/T_{c}\sim 4$ \citep[e.g.,][]{Maksimovic05,Tong19}, but in all simulations we assume $T_{h}/T_{c}=10$ to have larger initial growth rates and thereby reduce the saturation time of the instability. We perform two set of simulations. In the first set, $\beta_{c}=1$ and $\omega_{pe}/\Omega_{e}\approx 12.3$, while initial values of $u_{c}/v_{A}$ and $q_e/q_0$ are given in Table \ref{tab1} and indicated in Figure \ref{fig2}. In the second set, the initial electron heat flux is $q_{e}/q_0=0.45$ and simulations are performed at $\beta_{c}=0.4-3$. The corresponding initial values of $u_c/v_{A}$ and $\omega_{pe}/\Omega_{e}$ are given in Table \ref{tab2} and indicated in Figure \ref{fig2}. The realistic parameters in the solar wind are $\beta_{c}\sim 1$, $|u_{c}|/v_{A}\lesssim 10$ and $q_{e}/q_{0}\lesssim 1$ \citep[e.g.,][]{Tong18:arxiv,Tong19}. Several simulations of the first set are performed at $u_{c}/v_{A}$ larger than typically observed in the solar wind (Table \ref{tab1}). 

\begin{figure*}
    \centering
    \includegraphics[width=0.75\linewidth]{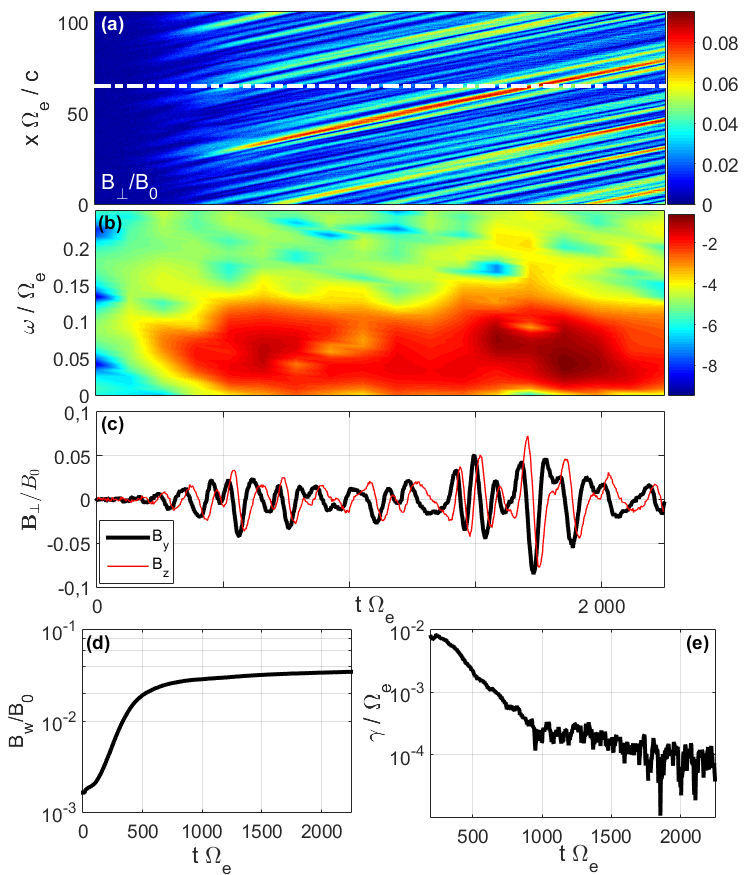}
    \caption{Results of one of the simulations of the first set ($\beta_{c}=1$, see Table \ref{tab1} for other initial parameters) with initial bulk velocity $u_c/v_A = -9$. Panel (a) presents evolution of the whistler wave magnetic field $B_\perp(x,t) = \left(B_y^2+B_z^2\right)^{1/2}$, white dashed line marks the position $x_0 = 65\,c/\Omega_e$, where we calculate the wave spectrum. Panels (b) presents the spectrum of the magnetic field component $B_y(x_0,t)$, while panel (c) shows temporal evolution of whistler wave magnetic fields $B_{y}(x_0,t)$ and $B_{z}(x_0,t)$, demonstrating $90^{\circ}$ phase-shift between $B_y$ and $B_z$. Panels (d,e) present the averaged (over the box) magnetic field amplitude $B_w(t) = \left(L_x^{-1}\int B^2_\perp(x,t) dx\right)^{1/2}$ and the growth rate $\gamma = B_w^{-1}dB_w/dt$.}
    \label{fig3}
\end{figure*}

Figure \ref{fig3} presents results of a simulation from the first set with $u_{c}=-9 \, v_{A}$. Panel (a) presents temporal evolution of the magnetic field $B_{\perp}(x,t) = \left(B_z^2+B_y^2\right)^{1/2}$ perpendicular to the background magnetic field, demonstrating the growth of electromagnetic waves propagating parallel to the electron heat flux. Panel (b) presents the spectrum of the magnetic field $B_{y}$ around the center of the simulation box, demonstrating that the waves are at the central frequency of about 0.05 $\Omega_e$ and have a bandwidth comparable to the central frequency. Panel (c) presents waveforms of $B_y$ and $B_z$ around the center of the simulation box and shows that the amplitude of the waves is less than 0.1 $B_0$. The 90$^{\circ}$ phase-shift between $B_y$ and $B_{z}$, meaning the right-hand polarization, and the wave propagation parallel the electron heat flux prove that the waves are the whistler waves produced by the WHFI. The central frequency of 0.05 $\Omega_{e}$ is consistent with the linear stability theory (Figure \ref{fig2}a). Panel (d) presents the temporal evolution of the whistler wave amplitude averaged over the simulation box, $B_{w}\equiv \left(L_{x}^{-1}\int B^2_{\perp}(x,t)\;dx\right)^{1/2}$. The instability saturates after about 2000 $\Omega_e^{-1}$ at the averaged whistler wave amplitudes of $0.03\,B_0$. Panel (e) presents the temporal evolution of the growth rate computed as $\gamma=B_{w}^{-1}dB_{w}/dt$ and shows that the initial growth rate $\gamma/\Omega_e\sim 0.01$ is consistent with the linear stability theory (Figure \ref{fig2}b).

\begin{figure*}
    \centering
    \includegraphics[width=0.95\linewidth]{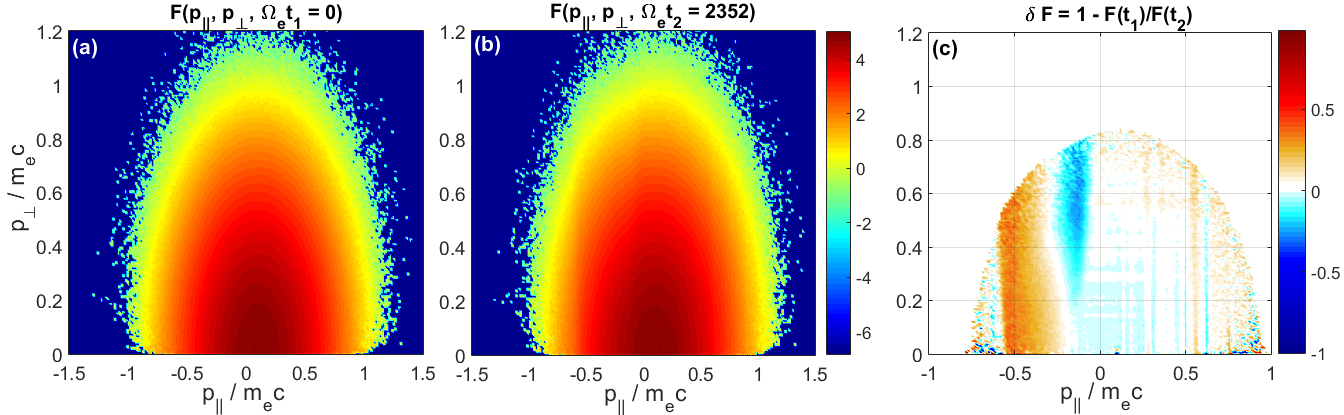}
    \caption{The electron VDFs in the simulation presented in Figure \ref{fig3}. Panels (a) and (b) present electron VDFs at the initial moment and after the saturation of the instability (after 2000 $\Omega_{e}^{-1}$), respectively. The VDFs are presented in $(p_{||},p_{\perp})$ plane, where $p_{||,\perp}$ are relativistic electron momenta. Panel (c) shows the difference between the VDFs, $\delta F(p_{||},p_{\perp})=1-F(p_{||},p_{\perp},t_1)/F(p_{||},p_{\perp},t_2)$. The phase space density of electrons is redistributed mainly in the region corresponding to the first normal cyclotron resonance, $v_{||}\approx (\omega-\Omega_{e})/k$. We masked out $\delta F(p_{||},p_{\perp})$ at $p_{||}^2+p_{\perp}^2\gtrsim m_{e}^2c^2$ to avoid effects of statistical noise due to small number of particle in these phase space regions.}
    \label{fig4}
\end{figure*}

Figure \ref{fig4} clarifies the mechanism of the saturation of the WHFI. Panel (a) presents the initial electron VDF, while panel (b) presents the electron VDF after 2000 $\Omega_e^{-1}$ that is after the saturation of the instability. Panel (c) presents the difference between electron VDFs at the initial moment and after the saturation of the instability. The scattering by the whistler waves results in redistribution of the phase space density of electrons with $v_{||}\sim -0.3\,c$. Namely, electrons with initial parallel velocities $v_{||}\gtrsim -0.3\,c$ are transported to parallel velocities $v_{||}\lesssim -0.3\,c$ and vice versa. These are actually electrons in the first normal cyclotron resonance, $v_{||}=v_{R}\approx (\omega-\Omega_{e})/k$ \citep[e.g.,][]{Shklyar09, SotoChavez2014}, because at $\omega_{pe}/\Omega_{e}\approx 9$, $\omega\sim 0.05\,\Omega_{e}$ and $kc/\omega_{pe}\approx 0.3$, we have $v_{R}\approx (\omega-\Omega_{e})/k\approx -0.3\,c$. The resonant electrons provide the energy for the whistler wave growth, while the gradual scattering of these electrons by the whistler waves results in saturation of the instability via formation of a ``plateau" on the electron VDF \citep[e.g.,][]{Kennel&Eng66,Sag&Gal69,Karpman74}. Panel (c) shows that the scattering is efficient for electrons with $|v_{||}-v_{R}|\lesssim \Delta v_{R}$, where $\Delta v_{R}\sim 0.2\,c$. The finite resonance width $\Delta v_{R}$ is due to the finite bandwidth of the whistler waves, $\Delta v_{R}\approx \Omega_{e}\Delta k/k^2$, where $\Delta k$ is the bandwidth of the whistler waves in the wavenumber space. Adopting the cold dispersion relation for estimates, $\omega\approx \Omega_{e}k^2c^2/\omega_{pe}^2$, we derive $\Delta v_{R}\approx |v_{R}|\;\Delta \omega/2\omega\sim 0.2\,c$, because the bandwidth $\Delta \omega$ is comparable to the central frequency $\omega$ (Figure \ref{fig3}b).

\begin{figure*}
    \centering
    \includegraphics[width=0.5\linewidth]{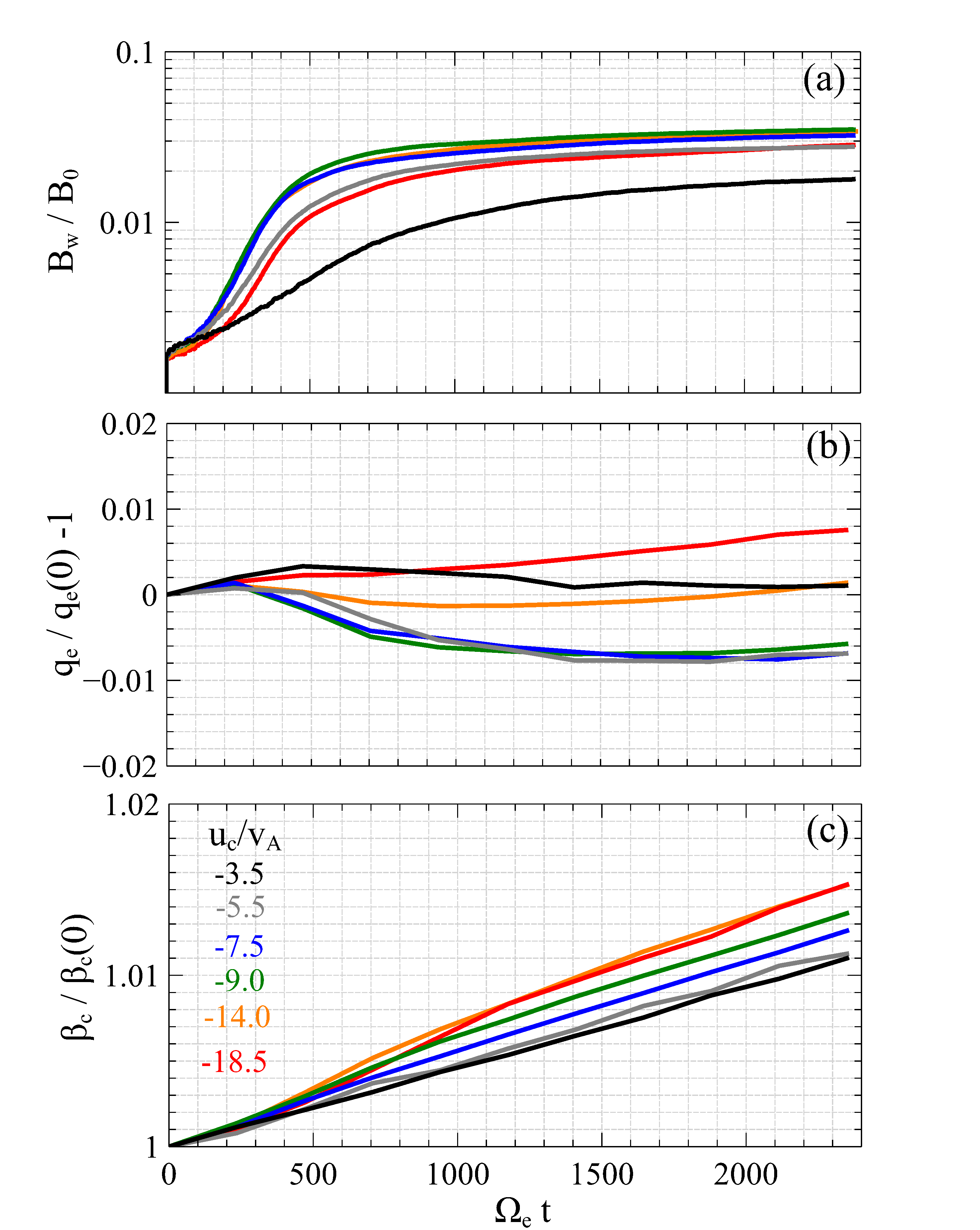}
    \caption{Results of the first set of simulations performed at $\beta_c = 1$ and various $u_c/v_A$ (Table \ref{tab1}). Panel (a) presents temporal evolution of the averaged whistler wave magnetic field amplitude $B_w/B_0$, panels (b) and (c) show temporal evolution of the electron heat flux $q_{e}$ and $\beta_{c}$ normalized to their initial values $q_{e}(0)$ and $\beta_{c}(0)$. Both $q_{e}$ and $\beta_{c}$ vary by less than a few percent in the course of the instability development and saturation. The monotonous growth of $\beta_{c}$ is due to a numerical heating of the core electron population.}
    \label{fig5}
\end{figure*}

Figure \ref{fig5} summarizes results of the first set of simulations performed at $\beta_{c}=1$ and various initial $u_c/v_{A}$ and $q_{e}/q_0$ (Table \ref{tab1}). Panel (a) shows the temporal evolution of the averaged whistler wave amplitudes and demonstrates that the amplitude in the saturation stage depends on the initial heat flux value. The whistler waves saturate at amplitudes of about $0.02 - 0.04\,B_0$. Panels (b) and (c) show that the scattering of electrons by the whistler waves result in variations of the electron heat flux and $\beta_{c}$ by less than a few percent. 
The variation of $\beta_{c}$ is mostly due to the numerical heating of cold electrons that is recognized by the monotonous growth of $\beta_{c}$. Though the initial heat flux values in the first set of simulations are in the range from 0.3 to 1.9 (Table \ref{tab1}), the heat flux values in the saturation stage differ from the initial values by less than a few percent. These results do not support the scenario that parallel whistler waves produced by the WHFI suppress the electron heat flux below the marginal stability threshold dependent on $\beta_{c}$ \citep{Gary1999b,Gary2000a}. According to that scenario, the heat flux values in the saturation stage would decrease down to approximately the same value in all the simulations, because the marginal stability threshold depends only on $\beta_{c}$. Yet, Figure \ref{fig5} shows that the heat flux does not decrease significantly, remaining almost constant for all the simulations in this set with different initial heat flux values (Table \ref{tab1}).

\begin{figure*}
    \centering
    \includegraphics[width=0.5\linewidth]{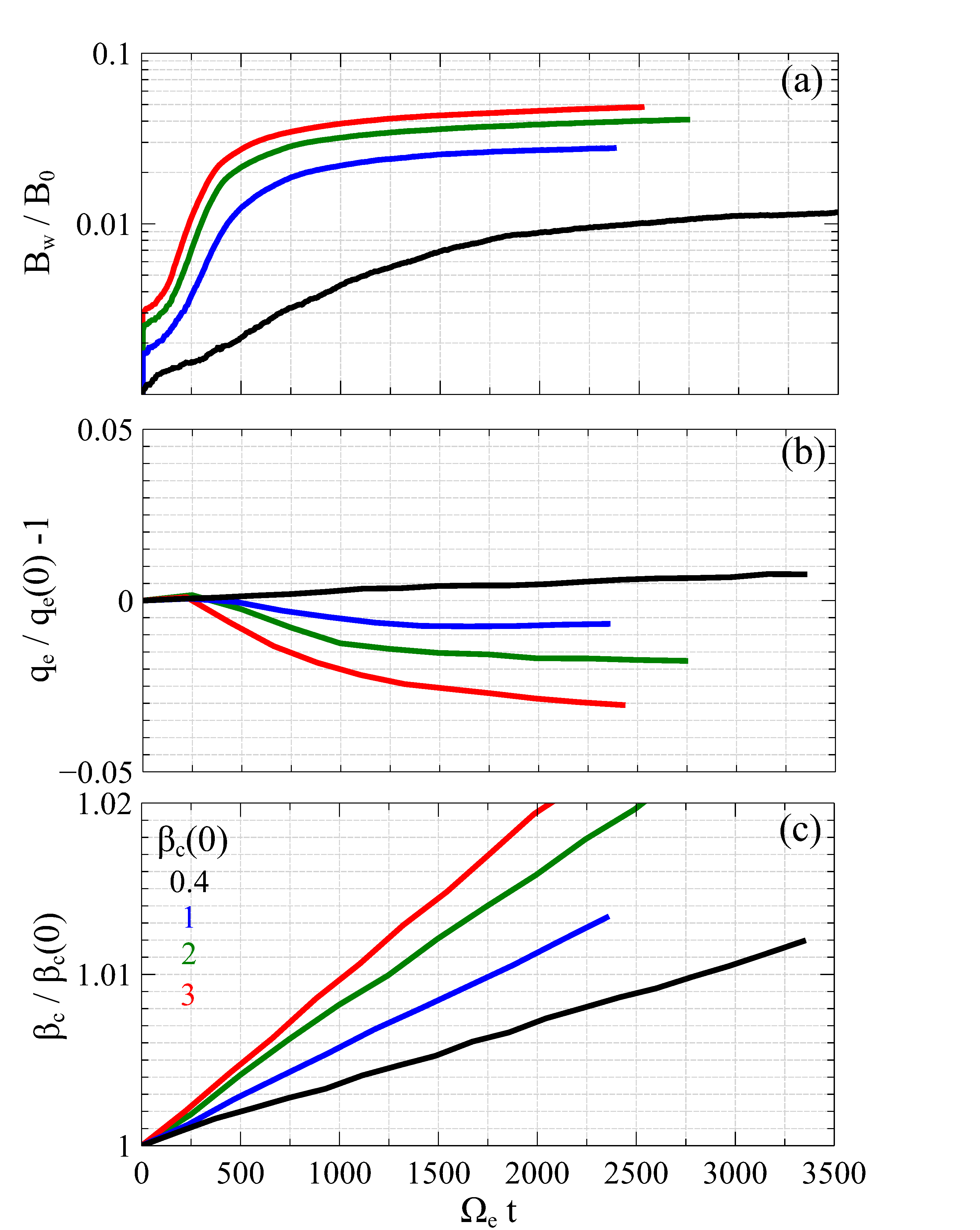}
    \caption{Results of the second set of simulations at various $\beta_{c}$ and fixed initial heat flux, $q_{e}/q_0=0.45$ (Table \ref{tab2}). Panel (a) presents temporal evolution of the averaged whistler wave magnetic field amplitude $B_w/B_0$, panels (b) and (c) show temporal evolution of the electron heat flux $q_{e}$ and $\beta_{c}$ normalized to their initial values $q_{e}(0)$ and $\beta_{c}(0)$. Both $q_{e}$ and $\beta_{c}$ vary by less than a few percent in the course of the instability development and saturation. The monotonous growth of $\beta_{c}$ is due to a numerical heating of the core electron population.}
    \label{fig6}
\end{figure*}

Figure \ref{fig6} summarizes results of the second set of simulations performed at the same initial heat flux $q_{e}/q_0=0.45$ and various $\beta_{c}$ (Table \ref{tab2}). Panel (a) presents the temporal evolution of the averaged whistler wave amplitude and demonstrates that the amplitude in the saturation stage depends on $\beta_{c}$. For $\beta_{c}=0.4 - 3$, the whistler waves saturate at amplitudes of $0.01-0.05\,B_0$. We should mention that in the simulation at $\beta_{c}=0.4$ the computations were performed up to $3500\,\Omega_{e}^{-1}$, because it takes longer for the instability to saturate at low $\beta_{c}$ due to smaller initial growth rates (Figure \ref{fig2}b). Panels (b) and (c) show that the scattering of electrons by the whistler waves results in variation of the heat flux and $\beta_{c}$ by less than a few percent. Again, these results do not support the scenario that parallel whistler waves produced by the WHFI suppress the electron heat flux below the marginal stability threshold \citep{Gary1999b,Gary2000a}. The heat flux values in the simulations with various $\beta_{c}$ differ by less than a few percent, while according to that scenario the heat flux values in the saturation stage would have been different, because the marginal stability threshold depends on $\beta_{c}$. 

\begin{figure*}
    \centering
    \includegraphics[width=1.0\linewidth]{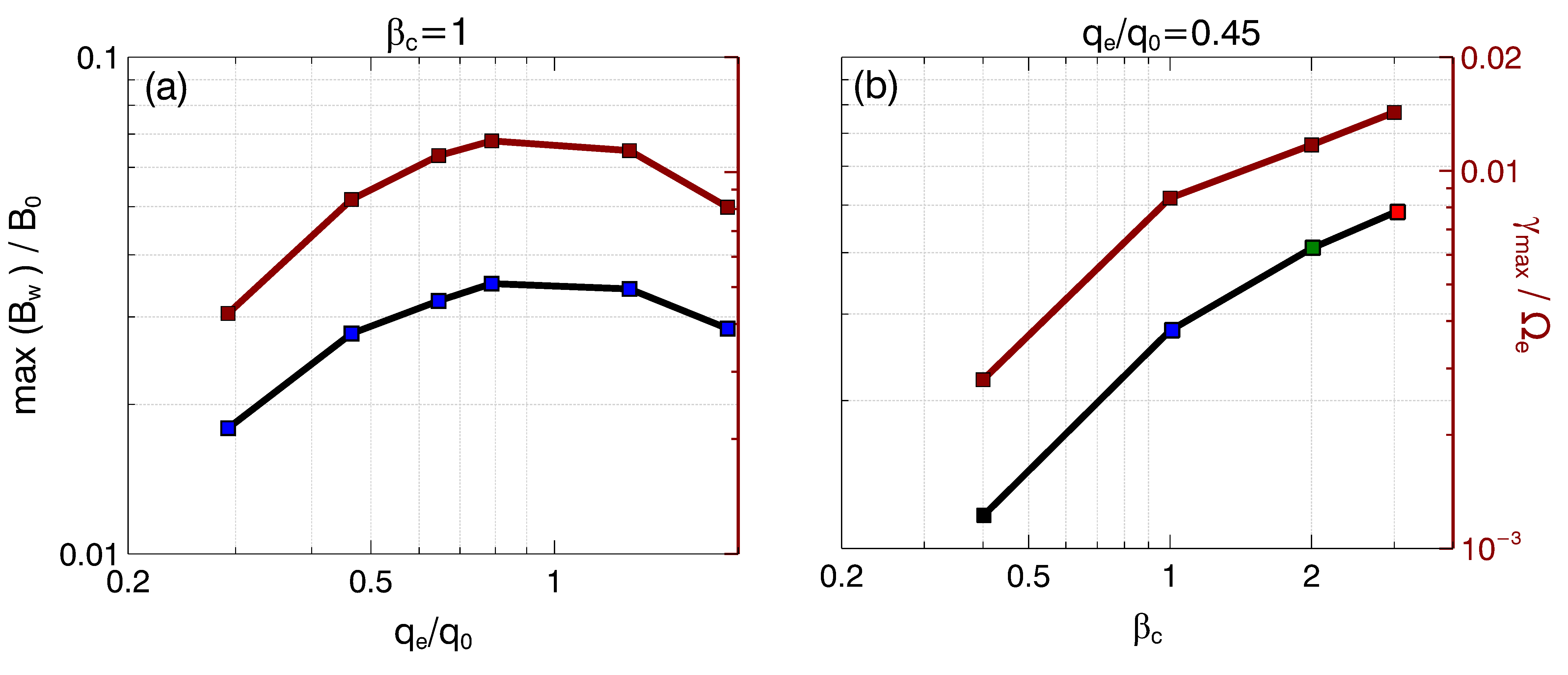}
    \caption{Panel (a) presents the saturated whistler wave amplitude versus initial normalized electron heat flux, as found in the first set of simulations performed at $\beta_{c}=1$ and various initial values of $u_c/v_{A}$ and $q_{e}/q_0$ (Table \ref{tab1}). Panel (b) presents the saturated whistler wave amplitude versus initial $\beta_{c}$, as found in the second set of simulations performed at $q_{e}/q_0=0.45$ and various initial values of $\beta_{c}$ (Table \ref{tab2}). The saturated whistler wave amplitudes ${\rm max}(B_{w})/B_0$ are determined after 2000 $\Omega_{e}^{-1}$ for $\beta_{c}\geq 1$ (see Figure \ref{fig5}) and after 3000 $\Omega_{e}^{-1}$ for $\beta_{c}=0.4$ (see Figure \ref{fig6}). In both panels we present the growth rates (brown curve) of the fastest growing whistler wave $\gamma_{\rm max}/\Omega_{e}$ derived from the linear stability theory (Figure \ref{fig2}). The squares in panels (a) and (b) are color coded as in Figure \ref{fig2}.}
    \label{fig7}
\end{figure*}

Figure \ref{fig7} presents correlations between the whistler wave amplitude in the saturation stage ${\rm max}(B_{w})/B_0$ and initial parameters of the electron VDF. Panel (a) presents ${\rm max}(B_{w})/B_0$ found in the first set of simulations versus initial $q_{e}/q_0$ and demonstrates that these quantities are positively correlated, though the correlation switches to negative at $q_{e}/q_0\gtrsim 0.8$. Panel (b) presents ${\rm max}(B_{w})/B_0$ found in the second set of simulations versus $\beta_{c}$ and demonstrates a positive correlation between these quantities. Panels (a) and (b) also present the linear growth rates $\gamma_{\rm max}/\Omega_{e}$ of the fastest growing whistler waves corresponding to the performed simulations (Figure \ref{fig2}). There is a clear positive correlation between the whistler wave amplitude in the saturation stage and the initial maximum growth rate. The results of the simulations are well fitted to 
\begin{eqnarray}
{\rm max}(B_{w})/B_0\sim A\; (\gamma_{\rm max}/\Omega_{e})^{\alpha}
\label{eq:1}
\end{eqnarray}
with $A\sim 0.57$ and $\alpha\sim 0.64$. The specific values of $A$ and $\alpha$ are likely dependent on $n_{c}/n_{e}$ and $T_{h}/T_{c}$ that were fixed in our simulations, but the positive correlation between ${\rm max}(B_{w})/B_0$ and $\gamma_{\rm max}/\Omega_{e}$ is expected to hold.

\section{Discussion \label{sec3}}

The electron heat flux is one of the leading terms in the electron energy balance in the expanding solar wind \citep[e.g.,][]{Cranmer09,Stverak15}, that is why kinetic processes regulating the electron heat flux are of fundamental importance in the solar wind physics. In this paper, we have presented results of 1D PIC simulations demonstrating that in a uniform plasma, the WHFI is incapable of regulating the electron heat flux. In other words, the presented simulations do not support the hypothesis proposed by \cite{Gary1999b} and \cite{,Gary2000a} that whistler waves produced by the WHFI suppress the electron heat flux below the $\beta_{e}-$dependent marginal stability threshold. Though that hypothesis was previously questioned based on theoretical estimates \citep{Pistinner1998a} and Ulysses measurements \citep{Scime01}, no detailed analysis of the nonlinear stage of the WHFI was provided to support those statements. We underline that our conclusions are based on the simulations restricted to parallel whistler waves. Though parallel whistler waves have the largest growth rates \citep{Gary75,Gary1994a}, 2D PIC simulations should be performed in the future to clarify effects of slightly oblique whistler waves produced by the WHFI.

The recent 1D PIC simulations by \cite{Roberg-Clark:2016} have also demonstrated that strictly parallel whistler waves are incapable of regulating the electron heat flux in a collisionless plasma. However, the initial electron VDF set by \cite{Roberg-Clark:2016} is an asymptotic solution of the collisional Fokker-Plank equation \citep[e.g.,][]{Pistinner1998a}, that is very different from electron VDFs typical for the solar wind \citep[][]{Feldman75,Maksimovic97,Tong19}. In contrast to \cite{Roberg-Clark:2016}, we have presented 1D PIC simulations of the classical WHFI with initial electron VDFs more relevant for the solar wind, so that the results of the simulations can be compared to in-situ measurements. We underline that in the realistic solar wind, the halo population is better fitted to $\kappa-$distribution with $\kappa\sim 4$, and typically $T_h/T_c\sim 4$ that is lower than used in our simulations \citep[e.g.,][]{Maksimovic05,Tong19}. Such a choice of distribution function and temperature ratio would affect the quantitative results for linear growth rates, and, consequently, the saturated wave amplitudes (see Eq. \ref{eq:1})), as we discuss below. But it does not affect the qualitative results for the WHFI, still producing low-amplitude quasi-parallel whistler waves.

The presented simulations show that whistler waves saturate at amplitudes of about $0.01-0.05\, B_0$. These amplitudes are comparable to the largest amplitudes of whistler waves measured in the solar wind at 1 AU \citep[][]{Tong19b}. Namely, the spacecraft measurements showed that whistler waves amplitudes are generally below $0.02\,B_0$, while typical amplitudes are of about $3\cdot10^{-3}\,B_0$ \citep[][]{Tong19b}. In the simulations with smaller density and temperature of the halo population ($n_{c}/n_{e}>0.85$ and $T_{h}/T_{c}<10$), whistler waves would certainly saturate at amplitudes smaller than $0.01-0.05\,B_0$ that would be consistent with the typical amplitudes found in the spacecraft measurements. The simulations have shown that the saturated whistler wave amplitude is positively correlated with $\beta_{c}$ that is consistent with the correlation between the whistler wave amplitude and $\beta_{e}$ revealed in the spacecraft measurements \citep[][]{Tong19b}.

Our results demonstrate that the saturated whistler wave amplitude  ${\rm max}(B_{w})/B_0$ depends non-monotonously on $q_{e}/q_0$. It is related to the fact that the initial (linear) maximum growth rate $\gamma_{\rm max}/\Omega_{e}$ depends on $q_{e}/q_0$ non-monotonously (Figure \ref{fig2}), while ${\rm max}(B_{w})/B_0\sim A(\gamma_{\rm max}/\Omega_{e})^{\alpha}$. Interestingly, similar correlation between the whistler wave amplitude and $q_{e}/q_0$ was revealed in the spacecraft measurements, though the correlation switches from positive to negative at $q_{e}/q_0\sim 0.3$ \citep{Tong19b}. Figure \ref{fig2} shows that at $T_{h}/T_{c}\sim 4$ that is more typical in the solar wind \citep[e.g.,][]{Maksimovic05,Tong19}, $\gamma_{\rm max}/\Omega_{e}$ reaches maximum at $q_{e}/q_0\sim 0.5$. Therefore, in simulations with $T_{h}/T_{c}=4$ the correlation between ${\rm max}(B_{w})/B_0$ and $q_{e}/q_0$ would switch from positive to negative at $q_{e}/q_0\sim 0.5$ that would be in better agreement with the spacecraft measurements. \cite{Tong19b} have considered whistler waves in the solar wind not disturbed by shocks, but we note that whistler waves produced by the WHFI were also identified around interplanetary shocks \citep{Wilson13,Liu18} and a positive correlation between the whistler wave amplitude and electron heat flux was reported \citep{Wilson13}.

The dependence of the saturated whistler wave amplitude on the initial maximum growth rate is similar to the one established by \cite{Tao17} in the quasi-linear analysis of the classical temperature-anisotropy instability \citep{Sagdeev60,Kennel66}. That is not surprising, because the nonlinear evolution of the WHFI seems to be quasi-linear in nature. Whistler waves have rather broad spectrum and saturate at relatively small amplitudes (Figure \ref{fig3}), so that the electron scattering is determined by the finite bandwidth, rather than by the finite amplitude of the whistler waves (Figure \ref{fig4}c). \cite{Tong19b} have recently demonstrated that the quasi-linear theory is indeed applicable to describe effects of whistler waves observed in the solar wind, thereby demonstrating an adequacy of the previous quasi-linear simulations of whistler wave effects in the solar wind \citep[e.g.,][]{Vocks05,Vocks12}.

The presented results indicate that there is a fundamental problem concerning processes controlling the electron heat conduction in the solar wind. We have shown that coherent parallel whistler waves observed in the solar wind \citep{Lacombe14,Stansby16,Kajdic16,Tong19b,Tong19} are unlikely to regulate the electron heat flux. We underline that we have addressed the evolution of the WHFI in a uniform plasma. Because the saturation time of the WHFI in physical units is from a few seconds to several minutes (see Figures \ref{fig5}a and \ref{fig6}a in this paper and Figure 16 in \cite{Tong19b}), whistler waves with typical group velocities of thousands km/s may cover a significant spatial region and can be influenced by low-frequency density and magnetic field fluctuations. The analysis of effects of low-frequency fluctuations on the nonlinear evolution of the WHFI deserves a separate study \citep[see, e.g.,][for analysis of non-uniform plasma density effects on a beam instability evolution]{Breizman70,Voshchepynets15}. In addition, the evolution of the WHFI may be influenced by the solar wind expansion, which provides the evolution of the background plasma parameters and may continuously support instability of a particular expanding plasma parcel \citep[e.g.,][for microscopic effects of the expansion]{Innocenti19}. The effects of a slightly oblique propagation \citep[within 20$^{\circ}$, see, e.g.,][]{Lacombe14,Kajdic16} of the observed whistler waves should be addressed in the future, though we note that the slight obliqueness may be also due to uncertainty of measurements. Incoherent whistler waves produced due to a turbulence cascade \citep[see, e.g.,][]{Gary12} could generally regulate the electron heat flux, but spacecraft measurements indicate that the magnetic field turbulence spectrum is likely dominated by kinetic Alfve\'n waves \citep[e.g.,][]{Bale05,Salem12,Chen13} and the contribution of whistler waves to the turbulence spectrum has not been estimated yet \citep[e.g.,][]{Narita16,Kellogg18}. Recent PIC simulations showed that rather oblique and large-amplitude electromagnetic whistler waves (wave normal angle of about $45^{\circ}$ and saturated amplitude $B_{w}\sim 0.3 B_0$) can be generated in a system with a sustained electron temperature gradient and suppress the electron heat flux \citep{komarov_2018,Roberg-Clark18:apj,Roberg18:prl}. The analysis of magnetic field spectra showed no evidence for such oblique whistler waves in the solar wind at 1 AU \citep[e.g.,][]{Tong19b}. The likely reason is that the electron VDF developing in those simulations is very different from what is observed in the solar wind. In particular, Landau resonant electrons damp too oblique whistler waves in the solar wind (that is why the WHFI of electron VDFs typical for the solar wind produces only quasi-parallel whistler waves) \citep{Gary1994a}, while Landau resonant electrons seem to drive oblique whistler waves in the simulations by \cite{Roberg18:prl}. Thus, it is currently not clear what processes regulate the electron heat flux in the solar wind \citep[see][for a recent discussion]{Vasko19,Verscharen19}, while simulations presented in this paper exclude coherent parallel whistler waves in a uniform plasma from the list of potential processes.

\section{conclusion \label{sec4}}

We have presented results of 1D PIC simulations of the whistler heat flux instability. We summarize the results as follows:
\begin{enumerate}
    \item In a uniform plasma, parallel whistler waves produced by the whistler heat flux instability are incapable of regulating the electron heat flux.
    \item Whistler waves saturate at amplitudes consistent with the recent spacecraft measurements. The simulations reproduce the correlations of the saturated whistler wave amplitude with the electron heat flux $q_{e}/q_0$ and $\beta_{e}$ revealed in the spacecraft measurements. 
    \item The saturated whistler wave amplitudes are dependent on the maximum growth rate computed from the linear theory, ${\rm max}(B_{w})/B_0\sim A(\gamma_{\rm max}/\Omega_{e})^{\alpha}$.
\end{enumerate}
The presented simulations exclude coherent parallel whistler waves observed in the solar wind from the list of potential processes regulating the electron heat flux. The effects of non-uniform background plasma density and magnetic field and slightly oblique propagation of the observed whistler waves remain to be addressed.

\acknowledgments
I. Kuzichev and A. R. Soto-Chavez were supported by the National Science Foundation (NSF) grant No. 1502923. They would like to acknowledge high-performance computing support from Cheyenne
(doi:10.5065/D6RX99HX) provided by NCAR's Computational and Information Systems Laboratory, sponsored by NSF. The NJIT Kong cluster at New Jersey Institute of Technology (Newark, NJ, USA) was also used. I. Kuzichev would also like to acknowledge the support of the \mbox{RBSPICE} Instrument project by JHU/APL subcontract 937836 to the New Jersey Institute of Technology under NASA Prime contract NAS5-01072. Y. Tong and S. D. Bale were supported in part by NASA contract NNN06AA01C.

\end{document}